\documentclass[twocolumn,prd,nofootinbib,floatfix,usenatbib]{revtex4}
\usepackage{graphicx,epsfig}
\usepackage{bm}

\newcommand{\beq}{\begin{equation}}
\newcommand{\eeq}{\end{equation}}
\newcommand{\fk}{\ensuremath{f_\mathrm{knee}}}

\newcommand{\simleq}{{\raise.0ex\hbox{$\mathchar"013C$}\mkern-14mu \lower1.2ex\hbox{$\mathchar"0218$}}}
\newcommand{\simgeq}{{\raise.0ex\hbox{$\mathchar"013E$}\mkern-14mu \lower1.2ex\hbox{$\mathchar"0218$}}}
\newcommand{\thcrit}{\ensuremath{\theta_\mathrm{c}}}
\newcommand{\thscan}{\ensuremath{\theta_\mathrm{s}}}
\newcommand{\thsci}{\ensuremath{\theta_{\mathrm{s} 1}}}
\newcommand{\thscii}{\ensuremath{\theta_{\mathrm{s} 2}}}
\newcommand{\vscan}{\ensuremath{v_\mathrm{s}}}
\newcommand{\dy}{\ensuremath{\Delta y}}
\newcommand{\traster}{\ensuremath{\tau_\mathrm{r}}}
\newcommand{\thk}{\ensuremath{\theta_k}}
\newcommand{\ngood}{\ensuremath{N_\mathrm{good}}}
\newcommand{\ntot}{\ensuremath{N_\mathrm{tot}}}
\newcommand{\nscan}{\ensuremath{N_\mathrm{s}}}
\newcommand{\sigsqw}{\ensuremath{\sigma^2_\mathrm{w}}}
\newcommand{\sigfourw}{\ensuremath{\sigma^4_\mathrm{w}}}
\newcommand{\dpsqr}{\ensuremath{\big [ \delta P(k) \big ]^2}}
\newcommand{\fsky}{\ensuremath{f_\mathrm{sky}}}

\newcommand{\dpnof}{\ensuremath{\delta P_{\mathrm{no }1/f}}}
\newcommand{\ApJS}{Astrophys. J. Supp. Ser.}
\newcommand{\AsAs}{Astron. Astrophys.}
\newcommand{\apjs}{\ApJS}
\newcommand{\aap}{\AsAs}

\begin{document}

\title{Power spectrum sensitivity of raster-scanned CMB experiments \\
in the presence of $1/f$ noise}
\author{Tom Crawford}
\email{tcrawfor@oddjob.uchicago.edu}
\affiliation{Department of Astronomy \& Astrophysics and Kavli Institute for Cosmological Physics, University of Chicago, Chicago, IL 60637}

\begin{abstract}
We investigate the effects of $1/f$ noise on the ability of a
particular class of Cosmic Microwave Background experiments to measure
the angular power spectrum of temperature anisotropy.  We concentrate
on experiments that operate primarily in raster-scan mode and develop
formalism that allows us to calculate analytically the effect of $1/f$
noise on power spectrum sensitivity for this class of experiments and
determine the benefits of raster-scanning at different angles relative
to the sky field versus scanning at only a single angle (cross-linking
versus not cross-linking).  We find that the sensitivity of such
experiments in the presence of $1/f$ noise is not significantly
degraded at moderate spatial scales ($\ell \sim 100$) for reasonable
values of scan speed and $1/f$ knee.  We further find that the
difference between cross-linked and non-cross-linked experiments is
small in all cases and that the non-cross-linked experiments are
preferred from a raw sensitivity standpoint in the noise-dominated
regime --- i.e., in experiments in which the instrument noise is
greater than the sample variance of the target power spectrum at the
scales of interest.  This analysis does {\it not} take into account
systematic effects.
\end{abstract}

\maketitle

\section{Introduction}
\label{section:introduction}
Scanning strategy is an important component in the planning of Cosmic
Microwave Background (CMB) anisotropy measurements, particularly in
the presence of long-timescale drifts in the gain or DC level of the
instrument response to incoming signal.  These drifts may be due to
intrinsic quantum processes in the detector or readout electronics,
temperature changes in the instrument environment, or slow changes in
the behavior of an external noise source, to name but a few examples.
It is common to lump all of these under the heading $1/f$ noise,
because the spectral density of such drifts often approximates a power
law in frequency.  The effect of such drifts on instrument sensitivity
to sky signals of interest depends on --- and often drives the design
of --- the instrument scan strategy.

It has become conventional wisdom in the field that sensitivity in the
presence of $1/f$ noise depends crucially on the degree of
cross-linking in the scan strategy --- that is, the number of
different scan angles from which each pixel in the map is observed
\cite[e.g.,][]{wright96,tegmark97c}.
Cross-linking is also often invoked as a means to minimize the 
effects of systematic effects such as scan-synchronous ground 
pickup and chopping mirror offsets; in this work, however, we 
only address cross-linking as a means to combat the effects of 
$1/f$ noise.  By the criterion of
degree of cross-linking, the
raster-scan strategy --- in which the instrument beam is scanned back
and forth across the sky field in azimuth and stepped in elevation ---
is maximally sub-optimal, in that the majority of map pixels are only
ever observed from one scan angle.  However, for many sub-orbital
(i.e., ground- and balloon-based) CMB missions, this is the preferred
mode of observing, because any elevation component to a scan will
result in a large scan-synchronous signal from the changing optical
depth of the atmosphere.  For most sub-orbital platforms, some degree
of cross-linking can be achieved with azimuth-only scans by observing
a sky field  at different times of day,
using sky rotation to change the scan angle.  For obvious reasons,
this technique does not work for instruments observing from the South
Pole.  
Despite this, successful measurements of CMB anisotropy have
been made from the South Pole using single-dish instruments in 
raster-scan mode \cite{runyan03a}, and instruments at the Pole 
that comprise a major part of the present and near future of 
CMB science are currently operating or plan to operate primarily 
in raster-scan mode. \cite{church03b,yoon06,ruhl04}

It seems therefore worthwhile to investigate the limits on CMB 
power spectrum 
sensitivity from a raster-scanned instrument in the presence of $1/f$
noise.
While the effects of $1/f$ noise on CMB 
power-spectrum sensitivity have been investigated by other authors, 
(e.g., \cite{stompor04}), 
the specific case of raster-scanning instruments has not been 
dealt with since the treatment of \cite{tegmark97c}, the conclusions 
of which are the focus of this work.
In particular, we wish to determine the relative merits of a scan
strategy with a single scan angle versus one 
in which the scan angle is changed on the timescale of the rotation 
of the Earth,
as these are the maximum amounts of cross-linking one can achieve 
with azimuth-only scans from a Polar and mid-latitude platform, 
respectively.  

To achieve this, we first develop formalism that enables us to
calculate analytically the effect of $1/f$ noise on power spectrum
sensitivity for raster-scanned experiments.  This effort is similar 
to that of 
\cite{wandelt03}, who performed the calculation for instruments that
scan along interleaved rings.  We apply the formalism we develop
to two fiducial
scan strategies, one with no cross-linking and one with the minimal
cross-linking available from a sub-orbital platform in the
mid-latitudes, using two parameterizations of $1/f$ noise: a toy model
which provides a simple instructive result and a more realistic model.
Finally, we discuss the results of these calculations in the context
of earlier work on the subject, particularly that of \cite{tegmark97c}.

\section{Formalism and toy-model results}
To try and get an analytical feel for the effects of $1/f$ noise on
CMB power spectrum estimation from raster-scanned observations of a
square sky field using a single-pixel instrument (the generalization
to an array instrument is trivial if the $1/f$ noise is uncorrelated
between array elements), we will begin by modeling $1/f$ noise as a
step function with one white-noise value above 
a particular frequency and 
another, much larger white-noise value below this frequency.  After
extracting the simple but illustrative result from this case, we will
apply the calculation to a slightly more realistic parameterization of
$1/f$ noise.
\subsection{Sensitivity from a single observation}
First, we note that for a small enough sky field, 
($\simleq 20^\circ$ in each direction),
we can use the flat-sky approximation to decompose the CMB fluctuations 
on this field in Fourier modes.  Following \cite{hivon02},
\beq
\label{eqn:flatsky1}
\Delta T(x,y)=\left ( 2 \pi \right )^{-2}
  \int d^2k \; a(\bm{k})e^{i(k_x x + k_y y)}.
\eeq
The $a(\bm{k})$ are zero-mean Gaussian variables with variance
\beq
\label{eqn:flatsky2}
\langle a(\bm{k}) \; a^*(\bm{k}') \rangle = P(k)\delta 
(\bm{k}-\bm{k'}),
\eeq
where $P(k) \simeq C_\ell$ and
$k = |\bm{k}| \simeq \ell$ in this approximation.

We wish to estimate $P(k)$ from our (noisy) map.  Following
\cite{knox95}, we note that for a set of $N$ independently measured
modes with $|\bm{k}|=k$ and variance $P(k) + P^{\mathrm{noise}}(k)$,
the variance on our estimate of $P(k)$ is:
\begin{eqnarray}
\label{eqn:knox}
\dpsqr &=& 
\langle \left (P^{\mathrm{est}}(k) - P(k) \right )^2 \rangle\\* 
\nonumber &=& \frac{2}{N}\left (P(k) + P^{\mathrm{noise}}(k) \right )^2.
\end{eqnarray}

Our sensitivity to the amplitude of an arbitrary spatial mode on the
sky $\Delta T(x_p,y_p) = a(m_p) \; m_p$ is given by:
\beq
\label{eqn:modesens}
\frac{1}{\sigma^2\left (a(m_p) \right)} = W_a = m^T_p W_{pp'} m_{p'},
\eeq
where $W_{pp'}$ is the pixel-pixel weight matrix of our 
map.\footnote{This is only strictly true for modes that are perfectly 
resolved by the instrument beam.}
If we have made a minimum-variance map from our observations, 
then 
\beq
\label{eqn:minvar}
W_{pp'} = A^T_{tp} W_{tt'} A_{t'p'},
\eeq
where $A_{tp}$ is the pointing matrix (the operator that deprojects 
a spatial mode into the timestream), and $W_{tt'}$ is the inverse of 
the timestream noise correlation matrix 
\beq
W_{tt'}=\langle n(t) n^T(t') \rangle ^{-1}.  
\eeq
If we have $1/f$ noise 
in our timestream, then $W_{tt'}$ will not be diagonal, but if the 
properties of the noise are stationary in time, then $W_{tt'}$ is 
circulant, and its Fourier transform 
\begin{eqnarray}
\tilde{W}_{ff'} &=& F_{ft} W_{tt'} F^T_{f't'} \\*
\nonumber &=& \frac{1}{P^{\mathrm{noise}}(f)} \ \delta_{ff'}
\end{eqnarray}
is diagonal.  We can then rewrite equation \ref{eqn:modesens} using
equation \ref{eqn:minvar} and inserting two pairs of forward and 
inverse Fourier transforms:
\label{eqn:fmodesens}
\begin{eqnarray}
W_a &=& m^T_p \left (A^T_{tp} W_{tt'} A_{t'p'} \right ) m_{p'} \\*
\nonumber &=& \left (A_{tp} m_p \right )^T F^T F W_{tt'} F^T F 
\left (A_{t'p'} m_{p'} \right ) \\*
\nonumber &=& \sum_f 
\frac{|\left [F A m \right ](f)|^2}{P^{\mathrm{noise}}(f)}.
\end{eqnarray}
where $\left [F A m \right ](f)$ is the Fourier coefficient at temporal 
frequency $f$ of the time-domain function that results from
deprojecting mode $m$ into the timestream.  (In other words, we have 
derived the unsurprising result that if the timestream noise 
is uncorrelated in the Fourier domain, our sensitivity to a mode 
on the sky is the sum of the (squared) 
Fourier coefficients of its deprojection 
into the timestream, weighted by the inverse Fourier-domain variance.)

For a raster-scan observation beginning at the origin of our sky
field, scanning at angular speed \vscan \ at an angle \thscan \ from
the $x$ direction of our field, and stepping a distance \dy \ in the
$y$ direction of our field every \traster \ seconds, a given Fourier
mode from the expansion in equation \ref{eqn:flatsky1} will deproject
into the timestream as:
\begin{eqnarray}
\label{eqn:tmodes}
A_{tp} \Delta T_p (\bm{k}) &=& a(\bm{k}) \times \\*
\nonumber && e^{i \left [ k_x \vscan \cos \thscan \; t + 
  k_y \left (\vscan \sin \thscan \; t + 
  \dy \; \mathrm{floor} (t / \traster) \right ) \right ] } \\*
\nonumber &\simeq& a(\bm{k}) 
e^{i \left [ k_x \vscan \cos \thscan + k_y 
\left (\vscan \sin \thscan + \dy / \traster \right ) \right ] t }.
\end{eqnarray}
where the ``floor'' function returns the nearest integer less
than or equal to its argument.  The second line of equation
\ref{eqn:tmodes} approximates the stepping in $y$ as a continuous (very
slow) scan in $y$.  It is clear that the deprojection of a single Fourier 
mode into the timestream of a raster-scanning experiment has power at 
only a single temporal frequency:
\beq
\label{eqn:fk1}
f(\bm{k}) \simeq \left | \frac{k_x}{2 \pi} \vscan \cos \thscan + 
\frac{k_y}{2 \pi} \left (\vscan \sin \thscan + \frac{\dy}{\traster} 
\right ) \right |.
\eeq

In the absence of mode-mode correlations, the 
variance on our measurement of the coefficient 
$a(\bm{k})$ 
is simply proportional to the timestream noise variance at the 
frequency $f(\bm{k})$:
\begin{eqnarray}
\sigma^2\left(a(\bm{k}) \right) &=& \frac{1}{W_a(\bm{k})} \\*
\nonumber &=& \frac{1}{\nscan} \; 
P^\mathrm{noise}\left (f(\bm{k}) \right),
\end{eqnarray}
where \nscan \ is the number of times the instrument is scanned 
across the field between elevation steps.
As traditionally defined \cite[e.g.][]{stompor04}, 
noise with a $1/f$ component has a power
spectrum:
\beq
\label{eqn:onef}
P^\mathrm{noise}(f) = \sigsqw \left (1 + \frac{\fk}{f}
\right ),
\eeq
where \sigsqw \ is the variance in the high-frequency
``white'' part of the power spectrum, and \fk \ is the 
frequency at which $P^\mathrm{noise}(f) = 2 \sigsqw$.
If we approximate this as
\beq
\label{eqn:approxonef}
P^\mathrm{noise}(f) = 
\cases{\sigsqw,& if $f > \fk$; \cr
\infty, &otherwise; \cr}
\eeq
then the variance on our measurement of $a(\bm{k})$ is:
\beq
\sigma^2\left ( a(\bm{k}) \right ) =
\cases 
{\sigsqw/\nscan, & if $f(\bm{k}) > \fk$; \cr
  \infty, &otherwise, \cr}
\eeq
and the variance on our estimate of the CMB power spectrum $P(k)$ from 
this map will be
\beq
\label{eqn:knox2}
\dpsqr = \frac{2}{\ngood(k)}\left (P(k) + \sigsqw / \nscan \right )^2,
\eeq
where $\ngood(k)$ \ is the number of modes with 
$|\bm{k}|=k$
that satisfy $f(\bm{k}) > \fk$.

If we define \thk \ as the angle between the $x$ direction of 
our sky field and the direction along which a particular Fourier 
mode is oscillating, then
\begin{eqnarray}
k_x &=& |\bm{k}| \cos \thk \\*
\nonumber k_y &=& |\bm{k}| \sin \thk,
\end{eqnarray}
and our criterion for goodness can be expressed as:
\beq
\label{eqn:angcrit1}
k \left | \cos \left (\thk - \thscan \right ) + \sin \thk \frac{\Delta_y}{\vscan \traster} \right | > \frac{2 \pi \fk}{\vscan}
\eeq
For raster timescales \traster \ that are sufficiently long compared to 
$1/\fk$, we can neglect the raster term 
in equation \ref{eqn:angcrit1} 
and write:
\beq
\label{eqn:angcrit2}
\left | \cos \left (\thk - \thscan \right ) \right | > \frac{2 \pi \fk}{k \vscan}.
\eeq
This allows us to define a critical angle
\beq
\thcrit(k) = \cos^{-1} \left(\frac{2 \pi \fk}{k \vscan} \right), 
\; 0 \le \thcrit \le \frac{\pi}{2}.
\eeq
If $2 \pi \fk/(k \vscan) > 1$,
then there are no good modes
at that $k$ value, and $\thcrit(k) = 0$.
(Note that for observations at a single scan angle, the choice 
of coordinate system that defines $k_x$ and $k_y$ is arbitrary, in 
which case we can define them such that $\thscan=0^\circ$, and our 
criterion for goodness is then simply $k_x > 2 \pi \fk / \vscan$.)

The {\it fraction} of good modes with a given $|\bm{k}| = k$ in the sky
map can now be expressed as:
\beq
\frac{\ngood(k)}{\ntot(k)} = \frac{2}{\pi} \; \thcrit(k).
\eeq
\ntot, the number of 
independent modes within a bin of width $\Delta k$ centered on $k$ for
a map covering a fraction \fsky \ of the sky 
is given by:
\beq
\ntot(k) = 2 k \fsky \Delta k,
\eeq
which allows us to write the variance on our measurement of 
$P(k)$ from this observation of this patch of sky as:
\beq
\label{eqn:dpapprox}
\dpsqr \simeq \frac{2}{2 k \fsky \Delta k} \; \frac{\pi}{2 \thcrit(k)}
\left (P(k) + \sigsqw / \nscan \right )^2.
\eeq

\subsection{Combining multiple observations}
\label{section:combine}
We now investigate the relative improvements in sensitivity from 
reobserving our sky patch, either using a different scan angle 
or the same scan angle as the first observation.  We will refer 
to these strategies as ``minimally cross-linked'' and 
``non-cross-linked.''

As discussed in the Introduction, sub-orbital CMB missions can achieve
a degree of cross-linking with azimuth-only scans by observing a patch
of sky at different times during the day.  In particular, the minimally
cross-linked strategy described above is achievable by observing a sky
field at the same elevation twice a day: once as the field rises,
once as it sets.  The maximum difference in scan angles available to a
sub-orbital platform depends on the latitude of the observing
platform:
\beq
|\thsci - \thscii| \leq 90^\circ - |\mathrm{lat}|.
\eeq
As noted earlier, no cross-linking is possible with azimuth-only 
scans from a Polar platform.  

\begin{figure}[!tpb]
  \centering
  \includegraphics[width=0.48\columnwidth]{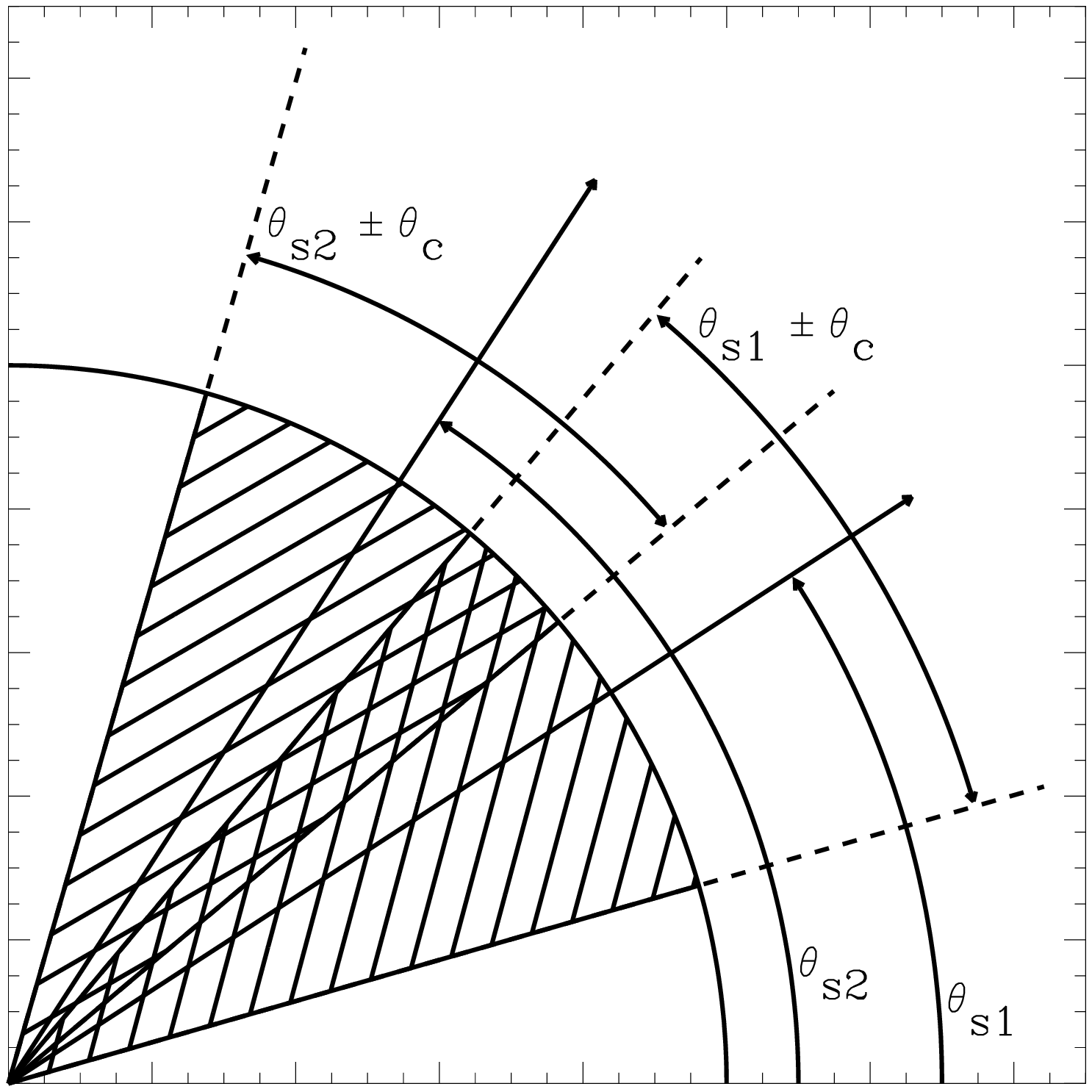}
  \includegraphics[width=0.48\columnwidth]{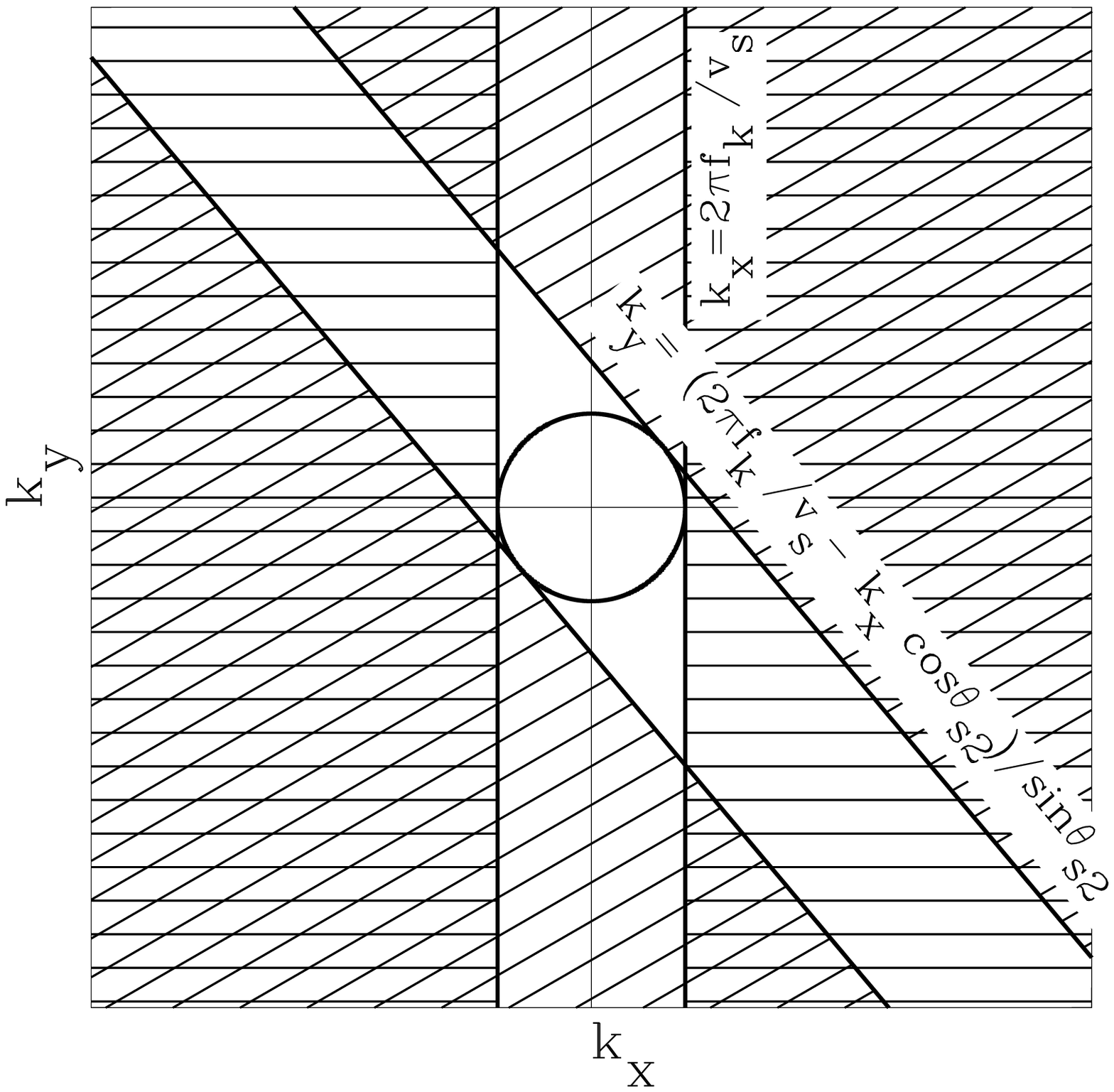}
  \caption{Graphical representation of the three classes of modes 
    enumerated in section \ref{section:combine}.  {\it Left Panel}: 
    Phasor plot of modes for a single $k=|\bm{k}|$.
    For two observations 
    at scan angles \thsci \ and \ \thscii, modes which fall within \thcrit \ 
    of either \thsci \ or \thscii \ but not both -- indicated by the 
    singly hatched regions -- will have variance $\sigsqw / \nscan$; 
    modes which 
    fall within \thcrit \ of both \thsci \ and \thscii \ -- indicated 
    by the doubly hatched region -- will have variance 
    $\sigsqw/2 \nscan$; modes
    which lie further than \thcrit \ from both \thsci \ and \thscii \ 
    (non-hatched regions) will have infinite variance.
    {\it Right Panel}: 2-d Fourier-space representation of these classes 
    of modes at all $k$ for $\thsci=0^\circ$ and $\thscii=40^\circ$.  The 
    circle at the center has radius $2 \pi \fk / \vscan$ and indicates 
    the modes that will have infinite variance no matter how many different 
    scan angles are used.
    \label{fig:modefrac}
  }
\end{figure}

If we reobserve our sky patch at the same scan angle
(no cross-linking), the fraction of
good modes will not change, but the noise variance in those modes will
decrease by a factor of 2.  If we reobserve the patch at a different
scan angle (minimal cross-linking), we will have three classes of modes:
\begin{enumerate}
\item{Modes which satisfy equation \ref{eqn:angcrit2} in both observations,}
\item{Modes which satisfy equation \ref{eqn:angcrit2} in one observation 
but not the other, and}
\item{Modes which satisfy equation \ref{eqn:angcrit2} in neither observation.}
\end{enumerate}
If we optimally combine the observations, these classes of modes will
have noise 
variance $\sigsqw/2 \nscan$, $\sigsqw / \nscan$, and $\infty$, 
respectively
(all modes will still have the same sample variance $P(k)$).  
If the two scan angles are sufficiently different 
($\left | \thsci - \thscii \right | > 2 \thcrit(k)$), the two 
observations will sample fully independent modes, and 
the fraction of modes falling into each noise variance category
can be expressed as:
\begin{eqnarray}
\label{eqn:modefrac1}
\frac{N_k(\sigma^2=\sigsqw/2 \nscan)}{\ntot(k)} &\equiv& 
\frac{N_1(k)}{\ntot(k)} = 0 \\*
\nonumber \frac{N_k(\sigma^2=\sigsqw / \nscan)}{\ntot(k)} &\equiv& 
\frac{N_2(k)}{\ntot(k)} = \frac{4 \thcrit(k)}{\pi} \\*
\nonumber \frac{N_k(\sigma^2=\infty)}{\ntot(k)} &\equiv& 
\frac{N_3(k)}{\ntot(k)} = 1 - \frac{4 \thcrit(k)}{\pi},
\end{eqnarray}
whereas if there is some overlap between modes sampled in the 
two observations (i.e., if 
$\left | \thsci - \thscii \right | \leq 2 \thcrit(k)$),
then the fraction of modes falling into each noise variance category
is:
\begin{eqnarray}
\label{eqn:modefrac2}
\frac{N_1(k)}{\ntot(k)} &=& 
\frac{2 \thcrit(k) - \left | \thsci - \thscii \right | }{\pi} \\* 
\nonumber \frac{N_2(k)}{\ntot(k)} &=& 
\frac{2 \left | \thsci - \thscii \right | }{\pi} \\* 
\nonumber \frac{N_3(k)}{\ntot(k)} &=& 
1 - \frac{2 \thcrit(k) + \left | \thsci - \thscii \right | }{\pi},
\end{eqnarray}
with the obvious limit that $0 \le N_i(k) / \ntot(k) \le 1$ in all cases and
that $\sum_i{N_i(k)} = \ntot(k)$.
Equations \ref{eqn:modefrac1} and \ref{eqn:modefrac2} 
are represented graphically in figure \ref{fig:modefrac}.

To get the minimum-variance estimate of the CMB power spectrum 
from these three classes of modes, we estimate $P(k)$ using 
each class individually and make a weighted mean of these estimates 
using inverse-variance weighting based on equation \ref{eqn:knox2}.  
In the noise-dominated regime
(where $\sigsqw / \nscan \gg P(k)$), 
the variance on our measurement of the CMB
power spectrum from these combined observations will be:
\begin{eqnarray}
\label{eqn:xlvnoxl} 
\dpsqr &=&
\frac{1}{\sum_{|\bm{k}_i| = k}
\big [ \delta P(\bm{k}_i) \big ]^{-2}} \\*
\nonumber &=& 
\left [ \frac{N_2(k)}{(\sigsqw / \nscan)^2} + 
\frac{N_1(k)}
{(\sigsqw/2 \nscan)^2} \right ]^{-1} \\*
\nonumber &=& 
\big [ \dpnof(k) \big ]^2 
\left [ \frac{N_2(k)}{4 \ntot(k)} + 
\; \frac{N_1(k)}{\ntot(k)} \right ]^{-1},
\end{eqnarray}
where we have identified the prefactor in equation \ref{eqn:xlvnoxl} 
\begin{eqnarray}
\label{eqn:whitenoise} 
\big [ \dpnof(k) \big ]^2 &=&
\frac{2}{\ntot(k)} \left [ P(k) +
\frac{\sigsqw}{2 \nscan} \right ]^2 \\*
\nonumber &=& \frac{\sigfourw}{4 \nscan^2 \; k \fsky \Delta k} 
\; \mathrm{ if } \;
\sigsqw / \nscan \gg P(k) 
\end{eqnarray}
as the variance our combined measurements would have 
if the instrument noise were equal to the high-frequency 
white-noise value at all temporal frequencies 
(i.e., without the $1/f$ noise component).  
The ratio of variance in the presence 
of our toy-model $1/f$ noise to the variance with no $1/f$
noise (still assuming $\sigsqw / \nscan \gg P(k)$) evaluates to
\beq
\label{eqn:indmodes}
\left [ \frac{\delta P(k)}{\dpnof(k)} \right ]^2 =
\frac{\pi}{\thcrit}
\eeq
in the independent-modes case 
(where $\left | \thsci - \thscii \right | > 2 \thcrit(k)$) and to
\beq
\label{eqn:overlap}
\left [ \frac{\delta P(k)}{\dpnof(k)} \right ]^2 =
\frac{\pi}{2 \thcrit - \left | \thscii-\thsci \right | / 2}
\eeq
in the overlapping-modes case 
(where $\left | \thsci - \thscii \right | \leq 2 \thcrit(k)$).
 
Equations \ref{eqn:indmodes} and \ref{eqn:overlap} contain the 
potentially surprising result
that in the noise-dominated regime, the best constraints on $P(k)$ are
obtained by reobserving the sky at the same scan angle (so that
$|\thscii-\thsci| = 0$).  That is to say, {\it the scan strategy 
without cross-linking outperforms the cross-linked strategy}.
It is trivial to show that in the
sample-variance-dominated regime (where $P(k) \gg \sigsqw / \nscan$), 
one comes
to the opposite conclusion, namely that the best constraints are
obtained in the minimally cross-linked strategy.
In fact, the question being considered here is very
similar to that of optimizing sky coverage vs.~observing time per
pixel in a CMB experiment.  In our case the question is how to divide
observing time between Fourier modes rather than spatial pixels, but
the optimization calculation is the same.

\subsection{Results for ``real'' $1/f$ noise}
\label{section:realonef}
The approximation in equation \ref{eqn:approxonef} leads to a 
concise, illustrative result, but there is no reason we cannot 
do the calculation for noise with a true $1/f$ spectrum (as defined 
in equation \ref{eqn:onef}).  For a single observation at 
scan angle \thscan, our best inverse-variance weighted estimate 
of $P(k)$ will have variance:
\begin{eqnarray}
\dpsqr &=& 
\frac{\pi}{2 k \fsky \Delta k} \times \\*
\nonumber && \Bigg \{ 
\int_0^{\pi/2} d(\thk - \thscan) \Bigg [P(k) + \\*
\nonumber && \frac{\sigsqw}{\nscan} \left (1 + \frac{2 \pi \fk}
{k \vscan \cos (\thk - \thscan)} \right) \Bigg ]^{-2} \Bigg \}^{-1}.
\end{eqnarray}
Optimally combining two observations at \thsci \ and \thscii \ 
yields a power spectrum measurement with variance:
\begin{eqnarray}
\label{eqn:realonef}
\dpsqr &=& 
\frac{2 \pi}{k \fsky \Delta k} 
\Bigg \{ 
\int_0^{2 \pi} d \thk \\*
\nonumber && \left [P(k) + 
\frac{\sigsqw}{\nscan} (1+q_1) || (1+q_2) \right ]^{-2} \Bigg \}^{-1},
\end{eqnarray}
where 
\beq
\label{eqn:qdef}
q_i= \frac{2 \pi \fk}{k \vscan \left | \cos(\thk - \theta_{s,i}) \right | } , 
\eeq
and the parallel operator is defined such that 
\beq
\label{eqn:pardef}
x_1 || x_2 = \left [ \frac{1}{x_1} + \frac{1}{x_2} \right ]^{-1}.
\eeq

\begin{figure}[tpbh]
  \centering
  \includegraphics[width=0.96\columnwidth]{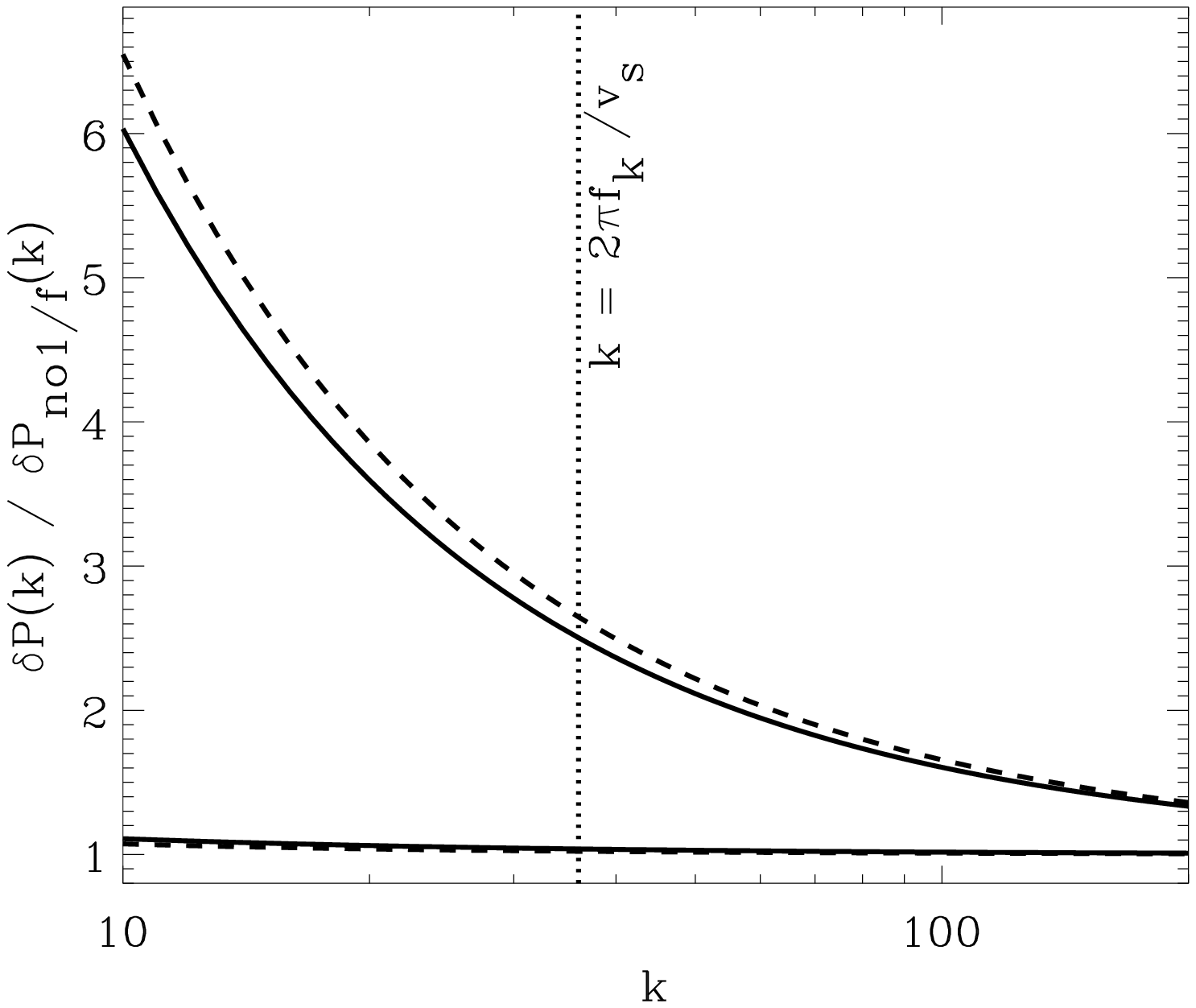}
  \caption{Ratio of $\delta P(k)$ to the no-$1/f$-noise value for 
    the non-cross-linked (solid line) 
    and minimally cross-linked (dashed line) scan strategies
    in two regimes of noise variance vs.~sample variance.
    {\it Upper curves:} $P(k) = \sigsqw/ \nscan/40$.
    {\it Lower curves:} $P(k) = 40 \ \sigsqw/ \nscan$. 
    In both cases, the observational parameters are: 
    $\fk = 0.1 \mathrm{Hz}$ and $\vscan = 1^\circ / \mathrm{s}$.
    In the minimally cross-linked cases, $\thscii=\thsci + 90^\circ$.
    \label{fig:mainresult}
  }
\end{figure}

Once again, the relative performance of the non-cross-linked and 
minimally cross-linked strategies depends on the relative size of 
the noise and sample variance.  Figure \ref{fig:mainresult} 
shows the ratio of $\delta P(k)$ to the value we would obtain 
with no $1/f$ noise 
for the two scan strategies 
and two noise- vs.~sample-variance regimes 
under consideration, using 
realistic values of \vscan \ and \fk.
%
Though the difference in the two strategies is less dramatic than the
factor of $\sqrt{2}$ in the toy model case 
(see equations \ref{eqn:indmodes} and \ref{eqn:overlap}), 
the behavior is qualitatively similar.  In the
noise-variance-dominated regime, the power spectrum sensitivity is 
dominated by the best-measured modes,
in which case it is advantageous to concentrate observing 
time on measuring a small number of modes very well; in the 
sample-variance-dominated regime, sensitivity is equal among all 
modes that are measured well enough to get below the sample 
variance, in which case it is better to spread the observing 
time around and measure as many modes as possible just 
well enough.  

\subsection{Results for other choices of scan parameters}
While the most important conclusion to be drawn from figure
\ref{fig:mainresult} is the relative performance between the
non-cross-linked and minimally cross-linked strategies, the  
amplitude of $\delta P(k)$ for both strategies compared to the case
with no $1/f$ noise is also of interest.  Not surprisingly, in 
the noise-variance-dominated regime, the value of $k$ at which 
the $1/f$ noise begins to dominate $\delta P(k)$ is very close 
to what would naively expect, namely
\beq
\label{eqn:konef}
k(1/f) = \frac{2 \pi \fk}{\vscan}.
\eeq
For the particular values of \fk \ and \vscan \ used in figure
\ref{fig:mainresult} ($\fk=0.1 \mathrm{Hz}$ and $\vscan = 1^\circ /
\mathrm{s}$), which are realistic approximations for the upcoming
generation of CMB experiments, $k(1/f) = 36$, indicating that for an
experiment which meets these criteria, $1/f$ noise will not limit
observations at least down to the $\ell \sim 100$ range (recall that
we are working in the flat-sky regime where $\ell \sim k$).  As for
experiments which have different values of \fk \ and \vscan, a quick
glance at equations \ref{eqn:realonef} - \ref{eqn:pardef} makes it
clear that for the noise-variance-dominated case, the results in
figure \ref{fig:mainresult} can be scaled from the values for \fk \
and \vscan \ used in the plot to arbitrary values by scaling the
x-axis using equation \ref{eqn:konef}.

The other key observing strategy design parameter for raster-scanned
experiments is the size and geometry of the sky patch observed in one
full raster.  The results in figure \ref{fig:mainresult} are
unaffected by changes in sky patch size and geometry (which only come
into equation \ref{eqn:realonef} in the \fsky \ term, which itself
drops out when we normalize to the no-$1/f$ case), except in that 
the extent of the largest dimension of the patch puts a fundamental 
lower limit on the $k$- (or $\ell$-) modes that can be measured.

\section{Discussion}
\subsection{Comparison with earlier work}
The results of equations \ref{eqn:indmodes} and \ref{eqn:overlap} and
figure \ref{fig:mainresult} seem in direct contradiction to the
conventional wisdom that cross-linking --- however minimal --- is
vital to CMB power spectrum sensitivity in the presence of $1/f$
noise.  In particular, \cite[T97]{tegmark97c} found an
order-of-magnitude difference in $P^\mathrm{noise}(k)$ for two scan
strategies very much like our non-cross-linked and minimally
cross-linked ones (cf.~T97's ``Serpentine'' and ``Fence'' scans).
However, the quantity plotted in T97 is effectively the noise power
averaged over all modes of a given $k$ (cf.~T97, equation 52), whereas
the total instrument sensitivity to signal at a given $k$ is
proportional to the sum of the weight, or the inverse noise power,
over all modes of that $k$.
The key to reconciling our result and
that of T97 is to understand that the excess noise power in the
non-cross-linked or Serpentine scan is concentrated in a few easily
identifiable bad modes, and that the remaining good modes are actually
measured better in this scan strategy than in the minimally
cross-linked, or Fence scan.  When the inverse noise power is summed
over all modes of a given $k$, it is the repeated measurement of the
good modes that wins out over measuring new modes 
--- at least in the
noise-variance-dominated regime.

\section{Conclusion}
We have calculated the sensitivity to the CMB power spectrum (or any
other angular power spectrum) for a raster-scanned instrument in the
presence of $1/f$ noise and compared this sensitivity in the
non-cross-linked and minimally cross-linked cases.  We have shown that
in the noise-variance-dominated regime, the non-cross-linked scan
strategy actually outperforms the minimally cross-linked scan
strategy.  From the viewpoint of noise sensitivity alone (i.e., not
taking into account systematic contaminants such as ground pickup or
chopping mirror offsets), this result indicates that CMB instruments
that are unable to easily cross-link scans are at no disadvantage.

\begin{acknowledgments}
The author would like to thank Tom Montroy, Bruce Winstein, Gil
Holder, Kendrick Smith, John Kovac, and an anonymous referee 
for useful discussions and
comments on early drafts.  This work was supported by NSF grant
No. OPP-0130612.
\end{acknowledgments}


\end{document}